\documentclass[pre,twocolumn]{revtex4}
\usepackage{amsmath}
\usepackage{graphicx}
\usepackage{latexsym}
\usepackage{amsfonts}
\usepackage{amssymb}
\usepackage{bm}
\usepackage{longtable}

\begin{document}

\title{Viscosity of Suspensions
of Hard and Soft Spheres
}

\author{George D. J. Phillies}
\email{phillies@wpi.edu}

\affiliation{Department of Physics, Worcester Polytechnic
Institute,Worcester, MA 01609}

\begin{abstract}
From a reanalysis of the published literature, 
the low-shear viscosity of suspensions of hard spheres is shown to have a dynamic crossover in its concentration dependence, from a stretched exponential at lower concentrations to a power law at elevated concentrations.  The crossover is sharp, with no transition region in which neither form applies, and occurs at a volume fraction (ca.\ 0.41) and relative viscosity (ca. 11) well below the sphere volume fraction and relative viscosity (0.494, 49, respectively) of the lower phase boundary of the hard sphere melting transition. For soft spheres -- taking many-arm star polymers as a model -- with increasing sphere hardness $\eta(\phi)$ shows a crossover from random-coil polymer 
behavior toward the behavior shown by true hard spheres.

\end{abstract}

\maketitle

\section{\label{sec:1} Introduction}

An interesting theme in modern physics is the flow of granular preparations.  While there is much interest in the flow of dry particulates (the 'sandpile' problem), an important issue in materials processing is the flow of 
solutions or suspensions of spherical or near-spherical particles.  Solution 
flows are subject to rheological constraints, the simplest being 
the low-shear viscosity $\eta$ of the solution.  $\eta$ 
depends very strongly on particle concentration.  If an accurate 
functional form for the concentration dependence of $\eta$ were available, 
one could predict $\eta$ at an arbitrary concentration via interpolation 
from a small number of precise measurements.   

Theoretical treatments of $\eta$ of hard sphere suspensions at elevated 
$\phi$ make a range of predictions.  Jones, et al.\cite{Jones} note results 
of Russell, et al.\cite{Russell}, but this treatment
underestimates $\eta$ for $\phi > 0.4$.  de Schepper, et 
al.\cite{deSchepper} show agreement between measurements of $\eta$ and a 
theoretical treatment of de Schepper, et al.\cite{Cohen}.  
However, Ref.\ \cite{Cohen}'s treatment ignores 
solvent-mediated hydrodynamic interactions, 
so its relevance to the dynamics of a hard-sphere suspension is 
not transparent.  Altenberger and Dahler\cite{Altenberger} 
use the positive-function renormalization 
group\cite{PFGR} to extrapolate $\eta$ from a low-concentration series, 
obtained from a hydrodynamic model, to 
elevated $\phi$.  With reasonable values for input parameters, the 
Altenberger-Dahler calculation works well for $\phi < 0.3$, but gives an $\eta$ that 
increases much too slowly with increasing volume fraction for $\phi > 0.5$.  

It was recently demonstrated in a short note \cite{phillies2002} for 
data\cite{Segre,Phan,Meeker} on polymethylmethacrylate spheres in {\em cis}-
decalin and other organic solvents that the viscosity $\eta$ of a hard sphere 
suspension has an accurate functional approximant.
At lower ($\phi < 0.42$) concentrations, one finds a stretched 
exponential 

\begin{equation} 
   \eta(\phi) = \eta_{o} \exp (\alpha \phi^{\nu}),
   \label{eq:stretched}
\end{equation}

\noindent while at higher ($\phi > 0.42$) concentrations a power law

\begin{equation}
   \eta(\phi) = \eta_{1} \phi^{x}.
   \label{eq:power}
\end{equation}

\noindent describes measurements well.  Here $\phi$ is the volume fraction of 
spheres, $\alpha$ is a scaling prefactor, $\nu$ and $x$ are scaling exponents, 
and $\eta_{o}$ and $\eta_{1}$ are prefactors with units of viscosity.  For hard 
spheres, $\phi = 1.0$ is unattainable; for hard spheres $\eta_{1}$ is therefore 
not the solution viscosity under physical circumstances.  As seen in ref.\ 
\cite{phillies2002}, the transition between these two forms is sharp, with no 
significant crossover regime in which neither form is valid.  This paper represents a substantial extension of the original note\cite{phillies2002}.

This paper explores the range of validity 
of eqs.\ \ref{eq:stretched} and \ref{eq:power}.  As shown below, these forms 
describe well the viscosity of a far wider range of systems and concentrations 
than reported in the preliminary note of ref.\ \cite{phillies2002}. 
Particles considered here extend in size upwards from micelles or protein molecules, with 
concentrations ranging from near-dilute to the random packing limit.  Some 
particles, such as silica nanospheres, are virtually incompressible; other 
particles are soft, permitting particle centers to approach to substantially 
less than twice the particle radius.  

\section{Background}

Interpretation of experimental data on hard sphere suspensions arises in the 
first instance from equilibrium statistical mechanics.  Close-packed hard 
spheres have a volume fraction $\phi_{cp} > 0.7$, but random close-packed 
systems generally attain volume fractions no greater than a nominal limit 
$\phi_{r} \approx 0.64$.  Hoover and Ree\cite{Hoover} interpret their computer 
simulations as implying (1) suspensions of neutral hard spheres having $\phi 
\leq \phi_{m} =0.494$ are single (meltlike) phase fluids, (2) hard spheres 
having  $\phi \geq \phi_{s} = 0.55$ occupy an expanded solidlike phase, and (3) 
at intermediate volume fractions $\phi_{m} \leq \phi \leq \phi_{s}$ the hard 
sphere system has an order-disorder transition, with an equilibrium between 
phases having volume fractions $\phi_{m}$ and $\phi_{s}$.  Hoover and Ree's 
interpretation requires that equilibrium hard sphere suspensions with $0.494 
\leq \phi \leq 0.55$ are biphasic.  

The concentration-dependent low-shear viscosity $\eta$ of hard sphere 
suspensions has been measured by a variety of authors, including Jones, et 
al.\cite{Jones},  Segre, et al.\cite{Segre}, Phan, et al.\cite{Phan},
Meeker, et al.\cite{Meeker}, Cheng and Schachman\cite{Cheng}, Marshall and Zukowski\cite{Marshall},  van der Werff and de Kruiff\cite{vanderWerff}, and de Kruif, et 
al.\cite{deKruif}.  These references are not in precise numerical 
agreement, perhaps because at large $\phi$ rheological properties are sensitive 
to small deviations of the suspended particles from perfect monodispersity and 
sphericity, and perhaps because small errors in determining $\phi$ lead to a 
large scatter in measured values of $\eta$ at a given nominal $\phi$.  Most of 
these results involve $\eta$ no more than a few hundred times the solvent 
viscosity $\eta_{s}$.  

Phenomenologically, at low concentrations the zero-shear viscosity 
of a hard sphere suspension can be described by a 
pseudovirial approximant
\begin{equation}
    \eta = \eta_{s}( 1 + k_{1} \phi + k_{2} \phi^{2} + k_{3} \phi^{3} 
    \ldots ) \label{eq:etalow} 
\end{equation}
where $k_{i}$ are expansion coefficients.  Cheng and Schachmann\cite{Cheng} 
confirmed the classic result $k_{1} = 2.5$ of Einstein for 260 nm diameter 
polystyrene spheres in 0.098 M NaCl.  They were unable to determine $k_{2}$ 
unequivocally.  

The viscosity of a hard-sphere solution increases markedly with increasing 
sphere concentration.  Data in refs.\ \cite{Segre}-\cite{Meeker} refer only to 
the meltlike phase $\phi \leq \phi_{m}$, but other references attained higher 
concentrations.  The recent and extremely thorough studies of Phan, et 
al.\cite{Phan} and Meeker, et al.\cite{Meeker}, both informed by the discussion 
between de Schepper, et al.\cite{deSchepper} and Segre, et al.\cite{Segrer}, 
find $\eta = 45 \pm 3$ and $\eta = 53 \pm 6$, respectively,  at volume fraction 
$\phi_{m} = 0.494$.  Marshall and Zukowski\cite{Marshall} report $\eta/\eta_{s} 
\approx 1 \cdot 10^{8}$ at $\phi \approx 0.6$.  

Roovers\cite{Roovers} provides extensive data on the viscosity of 
a model soft-sphere system, namely solutions of many-armed polybutadiene star 
polymers whose effective hardness is modulated by changing the number of 
arms.  The above references constitute a rheological phenomenology 
that can be compared with theoretical predictions.   

Interesting, related results that go beyond the scope of this paper include 
studies on the frequency-dependent linear viscoelastic behavior of hard sphere 
suspensions by van der Werff, et al.\cite{vanderWerff2} and on the rheology of 
zeroth-to-sixth generation dendrimers by Uppuluri, et al.\cite{Uppuluri}.  At 
elevated concentrations, shear thinning sets in at high shear rates 
$\dot{\gamma}$.   Marshall and Zukowski\cite{Marshall} report shear thinning 
for spheres of diameters 90, 210, and 286 nm above $\phi \approx 0.2$.  Jones, 
et al.\cite{Jones} record shear thinning for 50 nm spheres for $\phi > 0.3$.  
With 301 nm spheres, Segre, et al.\cite{Segre} observe shear thinning above 
$\phi \approx 0.4$.  There is thus no obvious correlation between sphere radius 
and the concentration for the onset of shear thinning, consistent with 
expectations for hard sphere systems.  Marshall and Zukowski\cite{Marshall} 
report that for $\phi > 0.5$ there is a marked increase in the characteristic 
time scale for shear thinning, while for $\phi > 0.56$ at elevated 
$\dot{\gamma}$ shear thinning is replaced by shear thickening.  However, shear 
thickening was not observed by Jones et al.\cite{Jones}.  Jones, et 
al.\cite{Jones} note that near the random-close-packed limiting concentration 
their systems gained a non-zero yield stress, with the elastic modulus 
satisfying $G'(\omega) \sim \omega^{0}$ at higher volume fractions.  

\section{Rationale}

The conjectured form for the concentration dependence of $\eta$ of a 
hard-sphere suspension, as successfully tested in ref.\ \cite{phillies2002}, 
arose from two sources.  First, the form works 
empirically\cite{Quinlan}-\cite{Phillies3} to high precision in 
certain other complex fluids involving the same fundamental forces but having 
non-spherical particles.  Second, an {\em ansatz} leading to the form and other 
results has been found on the basis of renormalization group 
concepts\cite{Phillies1999}.  

\begin{figure}

\includegraphics{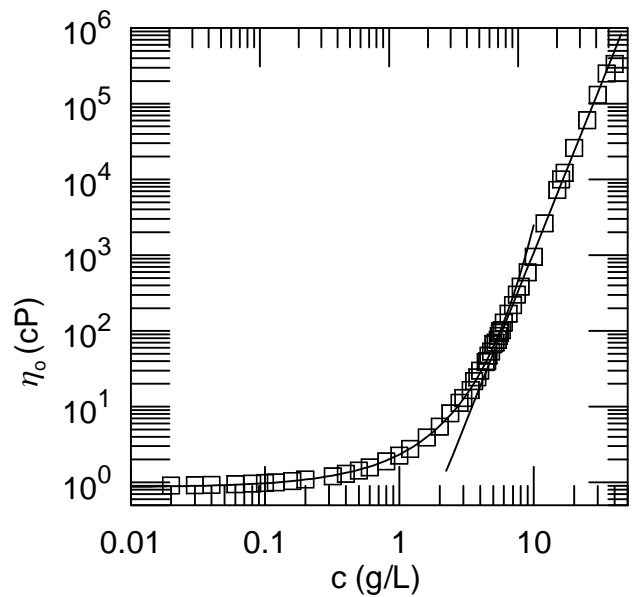} 

\caption{\label{figure1} Viscosity\cite{Quinlan} of 1MDa hydroxypropylcellulose: 
water as a function of concentration, and fits to a low-concentration stretched 
exponential regime (smooth curve, eq \ref{eq:stretched}) and a 
high-concentration power-law regime (straight line, eq \ref{eq:power}).
}
\end{figure}

Representative empirical data providing a rationale for the conjecture, namely 
measurements of Quinlan, et al.\cite{Quinlan} on $\eta$ of high-molecular-
weight hydroxypropylcellulose solutions, appear in Figure 1. The viscosity has 
a bifunctional concentration dependence, with two concentration regimes and a 
sharp crossover concentration.  At low concentrations, $\eta$ follows eq.\ 
\ref{eq:stretched}, while at higher concentrations the power law of eq.\ 
\ref{eq:power} is followed.  The crossover from stretched-exponential to 
power-law behavior occurs at a well-defined concentration $\phi^{+}$, with no 
indication near $\phi^{+}$ of a crossover regime separating the 
stretched-exponential and power-law regimes.  

This bifunctional concentration dependence of $\eta$ is not unique to 
hydroxypropylcellulose solutions.  Lin, et al.\cite{Lin} had previously 
reported that $\eta(\phi)$  in high-molecular-weight polyacrylic acid solutions 
had the same bifunctional concentration dependence.  Similar transitions have 
since been identified\cite{Peczak1,Phillies2,Phillies3} in data on some but not 
all solutions of high-molecular-weight polymers.  The crossover concentration 
$\phi^{+}$ is not a uniform multiple of the intrinsic viscosity $[\eta]$.  In 
different systems, $\phi^{+}$ variously appears\cite{Peczak1,Phillies2,Phillies3} in the range $4 \leq 
\phi^{+}[\eta] \leq 150$, or not at all.  

\begin{figure}

\includegraphics{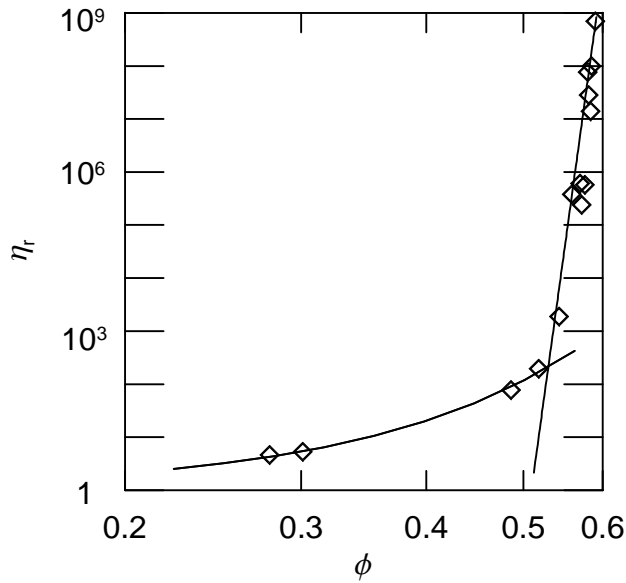} 

\caption{\label{figure2}  Viscosity of 205 nm sterically stabilized silica spheres, 
after ref \cite{Marshall}, and fits to eqs \ref{eq:stretched} (smooth curve) 
and \ref{eq:power} (straight line) using parameters in Table I.  
}
\end{figure}

\begin{figure}

\includegraphics{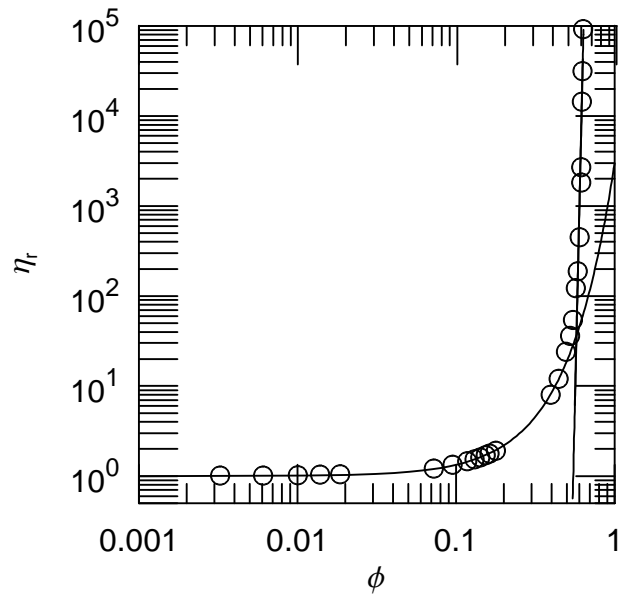} 

\caption{\label{figure3} Viscosity of 49 nm sterically stabilized silica spheres in 
Shellsol T, after ref \cite{Jones}. Other details as in Fig 2.  
}
\end{figure}  

This author\cite{Phillies1999} has previously advanced a rationale for the 
observed concentration dependence of $\eta$.  The rationale appears as  
part of a larger {\em ansatz} that correctly\cite{Phillies1999,Phillies1999b} 
predicts the frequency dependence of the loss and storage moduli of a polymer 
solution over a full range of frequencies.  Ref.\ \cite{Phillies1999}
proposed that 
the concentration dependence of $\eta$ out to high concentrations can be 
determined by applying the Altenberger-Dahler Positive Function Renormalization 
Group\cite{Altenberger} method to the known lower-concentration series 
expansion for $\eta$.  The functional form of $\eta(\phi)$ is determined 
by the dominant fixed point of the renormalization group at each concentration 
$\phi$, leading automatically to a low-concentration stretched-exponential concentration 
dependence arising from a fixed point at $\phi=0$ and a possible power-law 
concentration dependence at larger concentrations corresponding to a
large-concentration fixed point.  

Polymer solutions are, of course, not identical to solutions of spherical 
colloids.  However, the factor most likely to militate against the validity of 
the {\em ansatz} of ref.\ \cite{Phillies1999} for polymers is the hypothesized transition
from hydrodynamic-dominated dynamics to 
entanglement-dominated dynamics.  This transition might happen in polymer 
solutions, but suspended colloids cannot form entanglements.  The apparent 
success of the {\em ansatz} in polymer solutions does not prove that the {\em 
ansatz} will succeed in colloid solutions, but the most obvious reason for a 
failure of the {\em ansatz} refers to polymer systems, where the {\em ansatz} 
appears to succeed.  

\section{Hard Sphere Suspensions}

A literature search uncovered an extensive series of studies reported below.  Functional 
fits to the reported data were made to eqs.\ \ref{eq:stretched} and 
\ref{eq:power} using non-linear least squares based on the simplex algorithm.  
Most papers actually reported the reduced viscosity $\eta_{r} \equiv 
\eta/\eta_{s}$ rather than $\eta$ directly, so $\eta_{o} \approx 1$ 
often follows from fits to eq \ref{eq:stretched}.

Marshall and Zukowski studied\cite{Marshall} sterically stabilized 
silica spheres, diameters 82, 205, and 288 nm suspended in decalin, 
using a Couette double concentric cylinder geometry for sphere volume 
fractions up to 0.592 and viscosities $\eta/\eta_{s}$ up to almost 
$10^{9}$.  Marshall and Zukowski's results on the 82 and 288 nm spheres 
were confined to the high-concentration power-law regime; the 288 nm 
data have a single point below the power-law regime.  Figure 2 shows 
results on the 205 nm spheres, together with fits of the lower and upper 
concentration regions to eqs \ref{eq:stretched} and \ref{eq:power}, 
respectively.  Fitting parameters appear in Table I.   The proposed 
concentration dependences are seen from the Figure to describe each 
region well.  Because the power-law line is so steep, even very small 
experimental errors in the concentrations lead to the large 
RMS fractional error reported for the fit.

Jones, et al.\cite{Jones} measured the viscosity of 49-54 nm diameter silica 
spheres (different physical methods gave slightly different average diameters 
for the spheres) in Shellsol T (Shell Co.), using Ubbelohde capillary 
viscometers and three different cone and plate instruments for volume fractions 
up to 0.635 and relative viscosities as large as $9.2 \times 10^{4}$.  Figure 3 
presents their results and fits to our equations.  As seen in Figure 3, 
$\eta$ shows a stretched-exponential concentration dependence up to $\phi 
\approx 0.45$ with a fractional RMS error of 1.3\%, and a power-law 
concentration dependence for concentrations greater than approximately 0.55.  
In contrast to Figure 2, for $0.45 \leq \phi \leq 0.55$ a transition regime 
is apparent in Figure 3:  Several 
points do not quite lie on the lines describing the two functional forms. 

A qualitative difference between Figs.\ 1 and 3---and thus between polymer 
chains and hard spheres---is seen in the relative position of the 
stretched-exponential and power-law curves.  For a polymer solution (Figure 1) 
in the power-law regime the measured viscosities consistently lie below the 
viscosities predicted by extrapolating the stretched-exponential curve.  For a 
sphere suspension (Figure 3) in the power-law regime the measured viscosities 
lie above the extrapolated stretched-exponential curve.  In sphere 
suspensions, $\eta$ thus 
increases smoothly until the crossover.  Above the 
crossover the viscosity increases suddenly, far more steeply than expected from 
$\eta(\phi)$ below the crossover.  For polymers (Figure 
1) the crossover between the stretched-exponential and power-law regimes is 
analytic (first derivative continuous).   For hard spheres (Figure 3) the first 
derivative is not obviously continuous through the crossover.  

\begin{figure}
\includegraphics{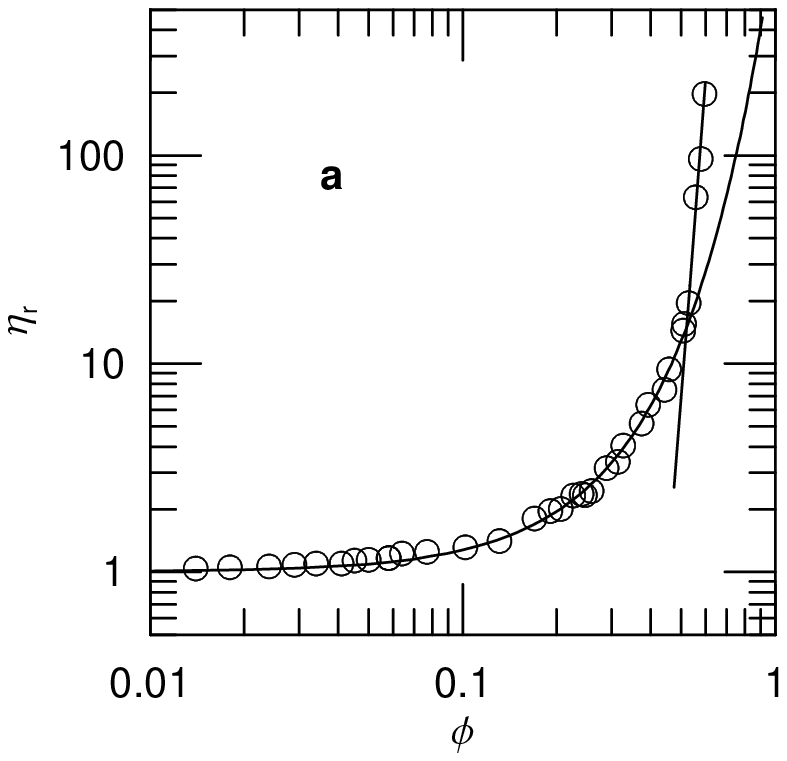} 
\includegraphics{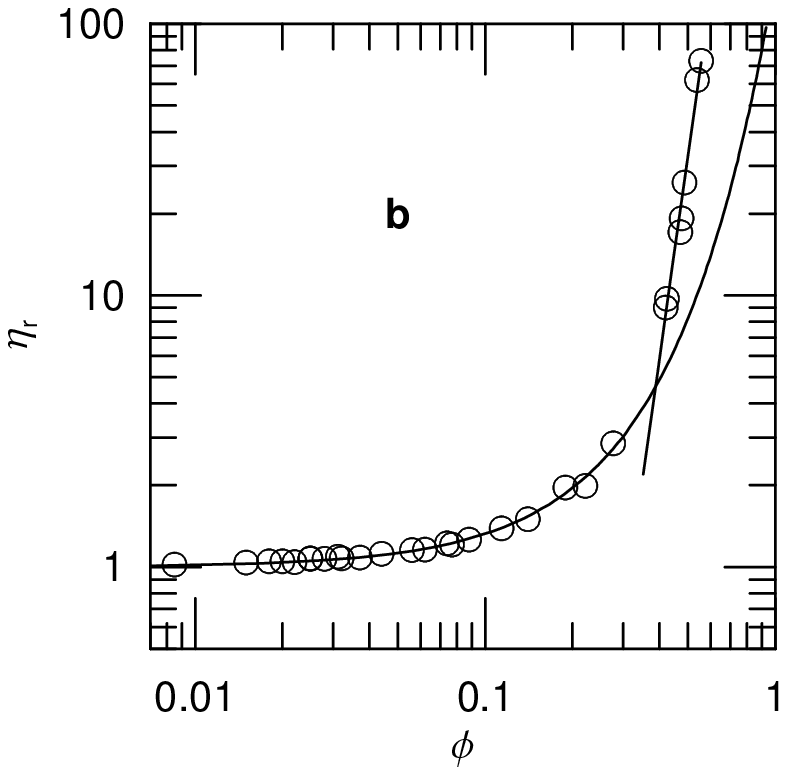}
\caption{\label{figure4} Viscosity of (a) 56nm and (b) 230 nm sterically stabilized 
silica spheres in decalin, after ref \cite{vanderWerff}. Other details as in 
Fig 2.  
}
\end{figure}

van der Werff, et al\cite{vanderWerff} measured steady-shear viscosities of 
monodisperse silica dispersions of diameters 56, 94, 153, and ca.\ 230 nm in 
cyclohexane, using an Ubbelohde capillary viscometer as well as a rheometer 
with Couette and parallel-plate measuring cells.  Figures 4a and 4b show their 
results for the smallest and largest spheres, together with the fitted curves.  
Table I includes fit parameters for all four sphere sizes.  RMS fractional
errors in the fits were 2-7\% for the stretched exponential regime, and in most 
cases 8-9\% for the power law regime.  In Fig 4a, data points appear almost 
exactly at the intersection of the two curves.  In these systems the crossover 
region between the two regimes, if any, must be extremely small.  

DeKruif, et al.\cite{deKruif} measured shear stress against shear rate 
for 156 nm silica spheres in cyclohexane at volume fractions $0.0006 
\leq \phi \leq 0.6$ and viscosities $\eta \leq 200 \eta_{s}$.  Their 
measurements were confined almost entirely to the  stretched-exponential
regime.  Fitting parameters appear in Table I.

\begin{figure}
\includegraphics{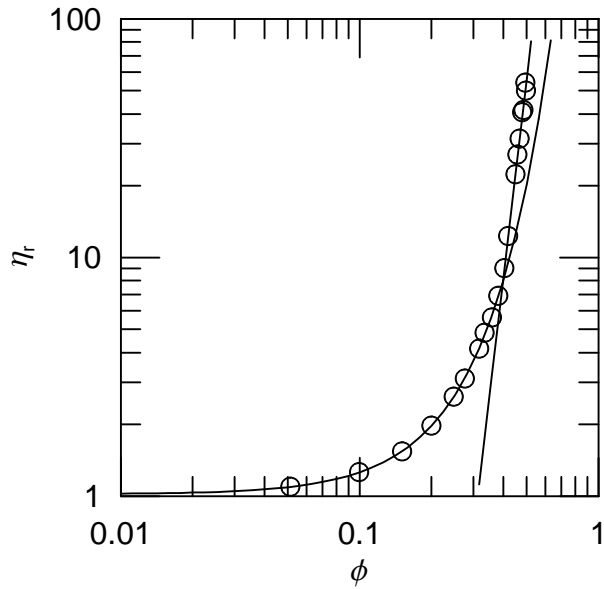} 
\caption{\label{figure7} Data of Segre, et al.\cite{Segre} on sterically-stabilized 602 nm polymethylmethacrylate spheres in cis-decalin. Other details as in 
Fig 2.  
}
\end{figure}

\begin{figure}
\includegraphics{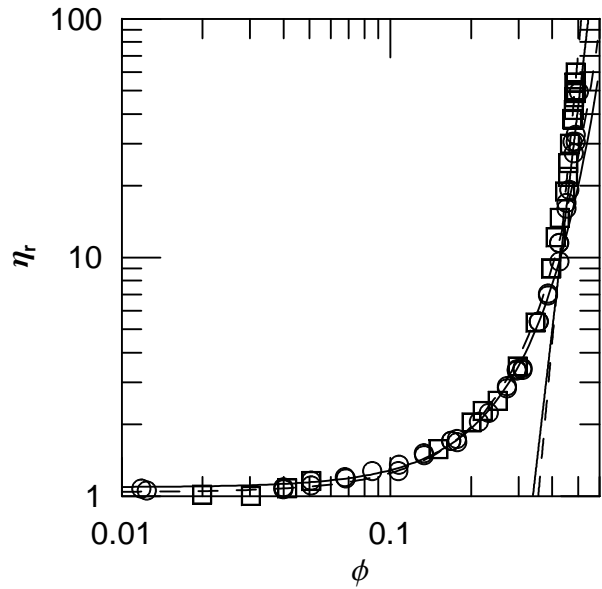}
\caption{\label{figure4}  Viscosity measurements of Phan, et al.\cite{Phan} (circles, solid lines; 580 and 640 nm diameter) and Meeker, et al.\cite{Meeker}(squares, dashed lines; 602 nm diameter) on polymethylmethacrylate spheres in various solvents. Other details as in 
Fig 2.  
}
\end{figure}

Segre et al.\cite{Segre} report the low-shear viscosity (obtained with a 
concentric-cylinder viscometer) and diffusion coefficient (from quasi-elastic 
light scattering) of 356 and 602 nm diameter polymethylmethacrylate spheres in 
{\em cis}-decalin.  Phan, et al.\cite{Phan} and Meeker, et al.\cite{Meeker} 
studied $\eta$ of 518 and 640 nm, and 602nm diameter, respectively, 
polymethylmethacrylate spheres in various solvents.  These data have previously been analyzed in 
ref. \cite{phillies2002}.  Fitting parameters are included in Table I. These papers, 
which confined themselves to concentrations $\phi \leq \phi_{m}$, found 
distinct stretched-exponential and power law regimes, the transition occurring 
at $0.41 \leq \phi^{+} \leq 0.43$ and  $\eta_{r} \approx 10-15$. The transition 
concentration $\phi^{+}$ is substantially less than $\phi_{m}$.  The 
differences between the sets of fitting parameters appears to reflect limits on 
measurement accuracy.  

\section{Soft Sphere Systems} 

The above results refer to spheres that are effectively non-deformable under 
the experimental conditions cited.  A comparison with properties of deformable 
spheres is allowed by the data of Roovers\cite{Roovers} on solutions of 32-, 
64-, 128-, and 270-arm polybutadiene star polymers.  Roovers\cite{Roovers} 
determined viscosities with Cannon-Ubbelohde viscometers, using multiple size 
viscometers and very long flow times, e.g., 1-2 hours, to confirm the absence 
of shear thinning.  Data were reported as a function of $\phi/\phi^{*}$, where 
$\phi^{*}$ is the overlap concentration defined to be
$\phi^{*} = 0.4/[\eta]$.

\begin{figure}

\includegraphics{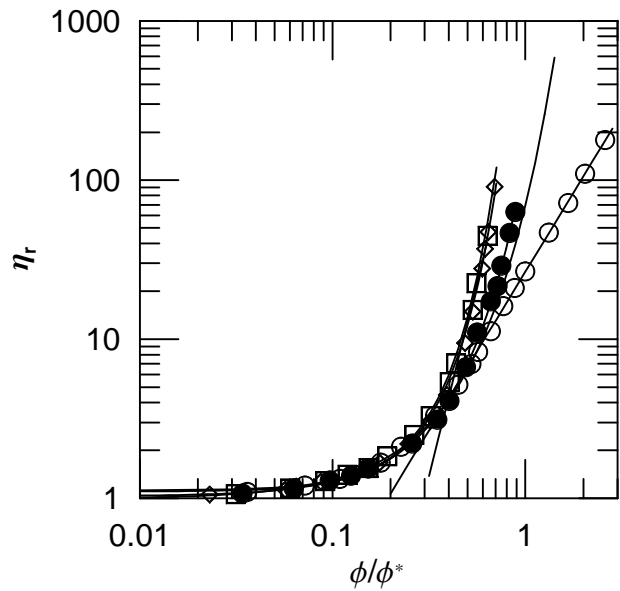} 

\caption{\label{figure5}  Viscosity \cite{Roovers} of $f$-armed polybutadiene star 
polymers in toluene, for $f$ of 32 ($\bigcirc$), 64 ($\bullet$), 128 
($\square$), and 270 ($\lozenge$), together with fits to eq 
\ref{eq:stretched} and (for the 32-arm stars) eq.\ \ref{eq:power}.  For the 32-
arm stars the power-law curve falls under the stretched-exponential curve.  For 
star polymers with $f > 32$ the power-law curve would be superposed on the 
displayed stretched-exponential curves. For hard spheres, the power-law curve 
lies above the stretched-exponential curve, as seen in Figures 1-5.
}
\end{figure}

Roovers' stars had a variety of arm lengths and molecular weights $0.4 \leq M 
\leq 11.2$ MDa. Figure 5 shows Roovers' data for ca.\ 3MDa stars; fit 
parameters appear in Table I.  Fits were also made to data on stars of the 
other molecular weights.  Roovers demonstrates that $\eta$ at fixed reduced 
concentration and arm number is independent of star molecular weight.  
Correspondingly, when eqs \ref{eq:stretched} and \ref{eq:power}, written as 
functions of $\phi/\phi^{*}$, are applied to Roovers' data, the resulting fit 
parameters are largely independent of molecular weight.  

In Figure 5, arm number $f$ increases from the lower-right-hand corner toward 
the upper-left-hand corner of the Figure.  For $\phi \leq 0.25 \phi^{*}$, 
$\eta$ is nearly independent of $f$.  At any larger concentration, $\eta$ of a 
32-arm star is less than $\eta$ of a 64-arm star, which is in turn less than 
$\eta$ of a 128-arm star.  At fixed $\phi/\phi^{*}$, increasing the number of 
arms from 128 to 270 has no further effect on $\eta$,  consistent with Roovers' 
interpretation that his materials have reached the limit of large $f$.  

\begin{figure}

\includegraphics{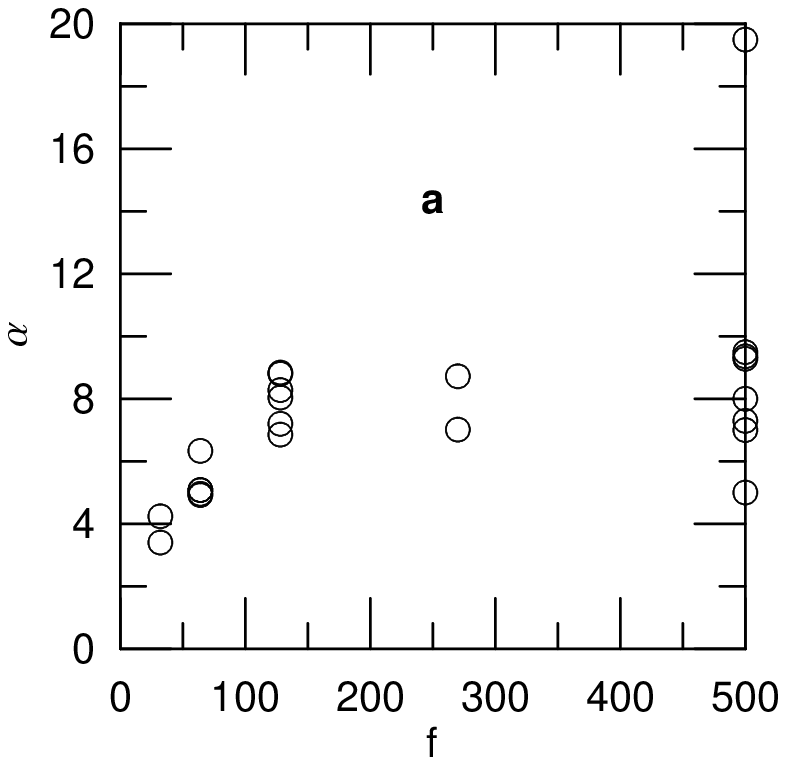} 

\includegraphics{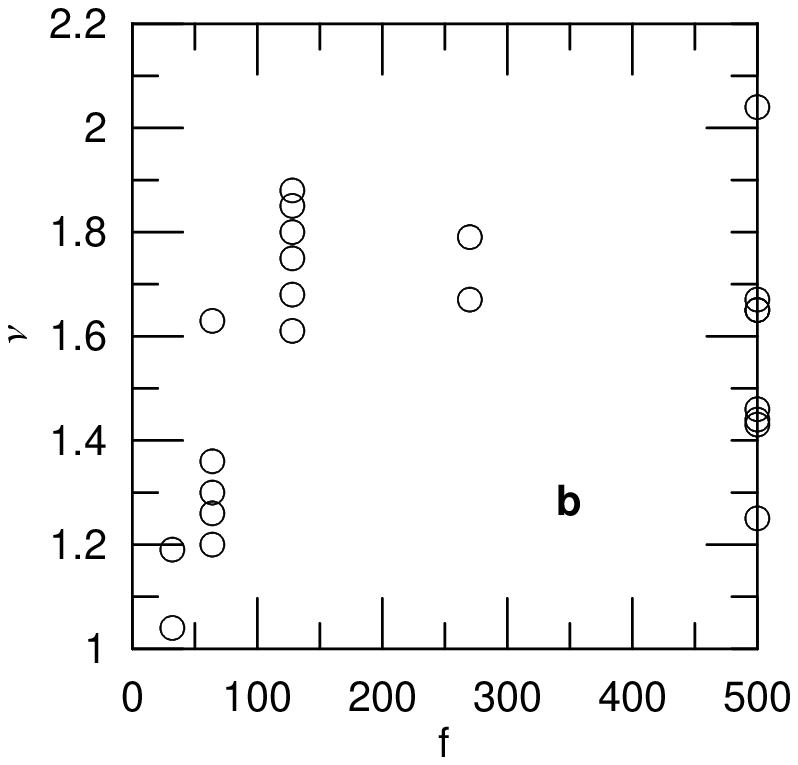} 

\caption{\label{figure6} a. $\alpha$ and b. $\nu$ as function of arm number $f$, 
based on fits to data of Roovers\cite{Roovers}, with hard sphere results (Table 
I) plotted at $f = 500$, showing $\eta(\phi)$ of 128- and 270-arm stars, as 
characterized by $\alpha$ and $\nu$, has reached the asymptotic hard-sphere 
limit.  
}
\end{figure}

In the lower concentration regime, $\alpha$ and $\nu$ describe $\eta(\phi)$.  
Figures 6a and 6b show $\alpha$ and $\nu$ as functions of arm number.  Points 
refer not only to the 3MDa stars of Fig.\ 5 but also to Roovers' other star 
polymers.  $\alpha$ nd $\nu$ for true hard spheres (Table I) are plotted in 
Fig.\ 6 as having a nominal arm number $f = 500$.   For $f \leq 128$, $\alpha$ 
and $\nu$ both increase with increasing $f$; for $f \geq 128$, $\alpha$ and 
$\nu$ are both approximately independent of $f$.  From Fig.\ 6, for $\phi < 
\phi^{+}$ not only do $f=128$ and $f = 270$ stars reach an asymptotic limit, 
but the limit is the hard-sphere limit.  

$\eta$ of a 32-arm star shows a clear large-$\phi$ power-law regime,
separated from the stretched-exponential regime via a crossover near $\phi
\approx 0.5$.  The crossover to this regime resembles that seen in Figure
1 for a polymer solution; the crossover is smooth and the measured $\eta$
in the power-law regime is less than the $\eta$ extrapolated to larger
$\phi$ from the stretched-exponential regime. If Roovers' 128- and 270-arm
polymers had reached their hard-sphere limit in their viscometric
properties, they would show a stretched-exponential to power-law
transition near $\phi/\phi^{*} \approx 0.4$.  Indeed, fits for these stars
for $\phi/\phi^{*} > 0.5$ to a power law obtain exponents $x$ of 5.9 and
6.8. These exponents are slightly less than the exponents of 8-12 typical
of true hard spheres, but are much larger than the $x \approx 2$ found for
32-arm stars.

\begin{table*}
\caption{ \label{table4} Fitting Parameters for Rheological Data.
Parameters from fits of $\eta_{o}$ or $\eta_{o}/\eta_{s}$ to 
eqs \ref{eq:stretched} and \ref{eq:power}, root-mean-square fractional errors 
expressed as a percent (\%R) in those fits, and crossover 
concentration $\phi^{+}$ and crossover viscosity $\eta^{+}$ between those 
forms.  Note that $\bar{\eta}$ represents the extended extrapolation of 
$\eta(\phi)$ to $\phi =1$; even a modest error in the slope 
$x$ leads to large 
errors in the estimated $\bar{\eta}$.  For the data of 
ref\ \cite{Roovers}, the first column gives the number of arms, not the
diameter of the star. 
}
\begin{ruledtabular} 
\begin{tabular}{|l|l|l|l|l|l|l|l|l|l|}
Diameter& $\eta_{s}$ & $\alpha$ & $\nu$ & \%R & $\bar{\eta}$ & $x$ & 
\%R & $\phi^{+}$ & $\eta^{+}$\\
\hline
288 nm\cite{Marshall}& - & - & - & - & $4 \times 10^{13}$ & 37.1 & 70 & 
- & - \\
205 nm\cite{Marshall}&1.0 & 19.5 & 2.04 &12 & $9.5 \times 10^{7}$& 10 & 
78 & 0.53 & 206 \\ 
49 nm \cite{Jones} & 1 & 8.0 & 1.44 & 1.3 
& $1 \times 10^{20}$   & 77 & 46 & 0.58 & 37.1 \\
56 nm \cite{vanderWerff} & 1 & 
7.0 & 1.46 & 5 & $4.7 \times 10^{6}$ & 19.4 & 8 & 0.52 & 14.8 \\
94  nm \cite{vanderWerff} & 1 & 9.5 & 1.67 & 7 & $1.7 \times 10^4$ & 8.97 
& 18 & 0.44 & 11.5  \\
153  nm \cite{vanderWerff} & 1 & 7.3 & 1.43 & 4 & $3.2 \times 10^{3}$ & 12.1 & 8 & 0.48 & 12.6 \\
230 nm \cite{vanderWerff}& 1  & 5 & 1.25 & 2 & $6.1 \times 10^{3}$ & 
7.57 & 9 & 0.39  & 4.6 \\
152 nm \cite{deKruif}& 1  & 9.27  & 1.65  & 6.4  & $6.1 \times 10^{8}$ &
12.4   & 2.0  & 0.49  & 18.3  \\
302nm \cite{Segre} & 1.02 & 9.35 & 1.65  & 0.9 & $7.5 \times 10^{6}$ & 
8.52 & 3.6&0.40  &7.9  \\ 
518 \&640nm \cite{Phan} & 1.09 & 9.98 & 1.78  & 3.6 & $3.0 \times 10^{4}$ & 
9.46 & 8.2 &0.42  & 10  \\ 
602nm \cite{Meeker} & 1.04 & 11.1 & 1.78  & 4.3 & $3.9 \times 10^{5}$ & 
12.5 & 8.2& 0.42  & 12  \\ 
$f=32$ \cite{Roovers} &0.99 & 4.24 & 1.19 & 1.8 &19.5 &1.96 &1.0 &- &- \\
$f=64$ \cite{Roovers} &1.03 & 4.91 & 1.36 & 4.5 &68.2 &3.9 &9.2 &- &- \\
$f=128$ \cite{Roovers} & 1.12 & 8.28 & 1.80 & 4.9 & 644 & 5.9 & 4.8 &- &- \\
$f=270$ \cite{Roovers} &1.09 & 7.01 & 1.67 & 3.8 &1030 &6.8 &7.4 &- &- \\
\end{tabular}
\end{ruledtabular}
\end{table*}

\section{Discussion} 

In the above, the published literature on $\eta$ of solutions of hard 
spheres was  re-examined.  As seen in Table I, in almost all cases eqs 
\ref{eq:stretched} and \ref{eq:power} describe $\eta(\phi)$ very well.  In each set of 
data, a transition in the functional form of $\eta$ is seen.  
For $\phi < \phi^{+}$, on a log-log plot $\eta(\phi)$ is a smooth curve 
with a continuously varying slope.  Above $\phi^{+}$, on the same log-log plot 
$\eta(\phi)$ appears as a straight line of large and unvarying slope.  

It has long been recognized that the viscosity of a hard-sphere suspension 
increases very sharply at elevated concentration.  In prior discussions it 
was not always noted that there is a qualitative change in 
the form of $\eta(\phi)$ when $\eta$ begins its sharp increase.  This 
apparent qualitative change in the functional form of $\eta(\phi)$ explains 
the observations of Jones, et al.\cite{Jones}, Russell, et al.\cite{Russell}, 
and Altenberger and Dahler\cite{Altenberger} that 
their low-volume-fraction forms 
for $\eta(\phi)$ work for $\phi < 0.3-0.4$, but fail badly 
when $\phi$ is taken to larger values.  

Root-mean-square fractional errors in the fits are generally in the range 1-
9\%.  For hard spheres, 
$\alpha$ is consistently in the range 7-11, while $\nu$ is almost always 
between 1.4 and 1.8.  The power-law exponent $x$ is most often around 10.  No 
pronounced dependence on any of these parameters on sphere radius is apparent, 
consistent with expectations that particle size should enter only through the 
sphere volume fraction $\phi$.  Deviations from the proposed forms are seen in 
the high-concentration data of Marshall and Zukowski\cite{Marshall} and Jones 
et al\cite{Jones}, in that their data scatters very substantially around the 
power law form.  

Table I reports the crossover concentrations $\phi^{+}$ and crossover
viscosities $\eta^{+}$ at which 
the stretched-exponential and power-law 
forms intersect. 
In most systems, $\phi^{+}$ is in the range 0.39-0.52, while $4.6 \leq 
\eta_{o}^{+} \leq 18.3$.

The transition in the 
concentration dependence of $\eta$ is not a dynamic simply reflection of 
the equilibrium phase boundary at $\phi_{m}$.  Phan, et al.\cite{Phan} and 
Meeker, et al.\cite{Meeker} did precise measurements of $\eta(\phi)$ up to 
$\phi_{m}$.  On analysis, their data shows that the dynamic transition occurs  
at $\phi^{+} \approx 0.42$, $\eta^{+} \approx 10-12$.  Thus, 
$\phi^{+}$ is well below the carefully determined $\phi_{m} = 0.494$, at which concentration
$\eta \approx 49$, well above $\eta^{+}$ that we have determined. 
In almost all systems $\eta^{+}$ is in the range 8-18, well below the 
experimental $\eta(\phi_{m})$.

While a systematic error in determining 
$\phi$ affects the determination of $\phi^{+}$, it has no effect on 
$\eta^{+}$.  Disagreements between determinations of $\eta^{+}$ thus cannot 
be related to difficulties in determining $\phi$.  An alternative plausible 
explanation for the experimental variation in $\eta^{+}$ is that the crossover 
is sensitive to details of the interparticle potential, so that varying 
degrees of sphere polydispersity or non-sphericity affect the crossover's 
location.

Several physical explanations for the apparent 
dynamic crossover suggest themselves:

1) The crossover might arise from a change in the static correlations 
in the system.  While it appears that $\phi^{+}$ is less than
the lower melting concentration $\phi_{m}$, 
dynamic effects are typically more sensitive to three-body 
correlations than are static correlations. One cannot readily  
exclude the  possibility that there are significant changes 
in $g^{(3)}({\bf r}_{1}, {\bf r}_{2}, {\bf r}_{3})$ of hard spheres,
as a precursor to the transition at $\phi = 
\phi_{m}$, at some 
concentration  such as $\phi^{+}$ that lies below $\phi_{m}$.

2) The crossover might result from additional dynamic correlations in the 
system at elevated $\phi$.  The oft-discussed formation of sphere doublets in 
shear at large $\phi$ would have this effect.  Such doublets could play the 
same qualitative role in sphere viscosity that is played in some treatments of 
polymer viscosity by the transient tube of the reptation model.  The doublets 
and the tube walls only appear at high concentration; each serves to obstruct 
lateral motion of the translating species.  

3) The crossover might be a purely mathematical consequence of taking 
low-concentration interactions to high concentration.  Altenberger and 
Dahler\cite{Altenberger,PFGR} have shown how a renormalization-group method can 
be used to calculate $\eta$ at elevated $\phi$.  Renormalization group 
methods involve series expansions around fixed points.  If a full 
renormalization group treatment of $\eta(\phi)$ were appropriate and if it 
had fixed points at $\phi=0$ and also at elevated $\phi$, then there would be a 
transition concentration at which one dominant fixed point replaces the other, 
for purely mathematical reasons.  Associated with this change in the identity 
of the fixed point would be a change in the functional dependence of 
$\eta$ on $\phi$.

\begin{acknowledgments}

The partial support of this work by the National Science
Foundation under Grant DMR99-85782 is gratefully acknowledged.

\end{acknowledgments}

\pagebreak

\end{document}